\documentclass[fleqn,10pt]{wlscirep}
\usepackage[T1]{fontenc}
\usepackage{hyperref}
\usepackage{caption}
\usepackage{subcaption}
\usepackage{multirow}
\DeclareMathOperator\erf{Q}
\newcommand{\E}{\mathrm{E}}
\newcommand{\Var}{\mathrm{Var}}
\title{Design and Evaluation of Crowdsourcing Platforms Based on Users’ Confidence Judgments}

\author[1]{Samin Nili Ahmadabadi}
\author[2]{Maryam Haghifam}
\author[1,*]{Vahid Shah-Mansouri}
\author[3,4]{Sara Ershadmanesh}
\affil[1]{School of Electrical and Computer Engineering, University of Tehran, College of Engineering, Tehran, Iran}
\affil[2]{Department of Computer Science, University of Toronto, Toronto, Canada}
\affil[3]{Department of Computational Neuroscience, MPI for Biological Cybernetics, Tuebingen, Germany}
\affil[4]{School of Cognitive Sciences, Institute for Research in Fundamental Sciences, Tehran, Iran}

\affil[*]{vmansouri@ut.ac.ir}


\begin{abstract}
Crowdsourcing deals with solving problems by assigning them to a large number of non-experts called crowd using their spare time. In these systems, the final answer to the question is determined by summing up the votes obtained from the community. The popularity of these systems has increased by facilitating access for community members through mobile phones and the Internet.\\
One of the issues raised in crowdsourcing is how to choose people and how to collect answers. Usually, users are separated based on their performance in a pre-test. Designing the pre-test for performance calculation is  challenging; The pre-test questions should be selected to assess characteristics in individuals that are relevant to the main questions.\\
One of the ways to increase the accuracy of crowdsourcing systems is by considering individuals' cognitive characteristics and decision-making models to form a crowd and improve the estimation of their answer accuracy to questions. People can estimate the correctness of their responses while making a decision. The accuracy of this estimate is determined by a quantity called metacognition ability. Metacoginition is referred to the case where the confidence level is considered along with the answer to increase the accuracy of the solution \cite{maniscalco2014signal}. In this paper, by both mathematical and experimental analysis, we would answer the following question: Is it possible to improve the performance of a crowdsourcing system by understanding individuals' metacognition and recording and utilizing users' confidence in their answers?
\end{abstract}
\begin{document}

\flushbottom
\maketitle

\thispagestyle{empty}

\section{Introduction}\label{Introduction}
The goal of crowdsourcing is to entrust answering questions, called tasks, to non-experts and using their idle time at a low cost. In these systems, cost is defined as the time consumed to decide and the payment to the users. Crowdsourcing tries to use the extra processing power of millions of human brains. The experience with open-source software proved that a network of enthusiastic volunteers and highly paid developers at Microsoft and other companies could develop software. Wikipedia showed that this model could be used to create a surprisingly distributed online encyclopedia \cite{howe2006rise}.\\

Given that most crowdsourcing systems are online, it usually does not matter where people are located. These systems help companies, industries, researchers, etc. in finding solutions to their problems or required data in a more affordable and faster way. Also, crowdsourcing systems can be an easy source of income for people in their spare time. Furthermore, people with disabilities can easily participate in these systems.\\

Many discussions in the field of crowdsourcing systems deal with solutions to avoid and minimize errors \cite{li2013error, sautter2013high}. In this regard, this research deals with the use of the decision-making model of people in such systems. Knowing how people make decisions makes it possible to predict the probability of the person answering the question correctly. One unsolved question in this area is whether it is possible to improve the system's performance by asking people's confidence in their answers and how this improvement can be achieved. In literature, metacognition ability is defined based on users' responses and confidence in order to see how accurately they can assess their performance. Thus, Metacognition ability is a predictor of how well users can respond and how well they are reporting their confidence \cite{fleming2017hmeta}. \\

In typical crowdsourcing systems, people are separated based on efficiency, and the majority vote algorithm is used to aggregate votes. In order to investigate the impact of metacognition and people's decision-making model in crowdsourcing systems, in this paper, we present a new method for screening people relying on metacognition and aggregation of votes with confidence and measuring their performance against a conventional system. Metacognition ability-based crowdsourcing systems can be useful in two main cases:
\begin{itemize}
	\item Metacognition ability is the same among different tasks \cite{veenman1997generality,mazancieux2018retrospective}. Therefore, this parameter can be measured once and recruit users for several tasks according to their metacognition ability.
	\item Sometimes, people choose an answer that are not completely sure about its correctness. Asking how confident they are may help the system to change the weight of their decision in the final aggregation.
\end{itemize}

In this research, our contributions  are as follows:
\begin{itemize}
	\item We propose a novel crowdsourcing system based on the users' reported confidence and their metacognition ability.
	\item With the help of simulation, we check the superiority of our proposed system for different communities. With the help of the results, we have identified the characteristics of the communities in which our system works better.
	\item By designing and implementing a real-world task, we prove our hypothesis that our system outperforms the typical crowdsourcing systems in certain communities. We ask people to identify the gender of the author of several tweets. This experiment's results show that using the proposed crowdsourcing system reduces the error.
\end{itemize}

In Section \ref{literature review}, we introduce the components of crowdsourcing systems and examine the solutions for various problems in these systems. Then we will introduce and review several practical and commercial examples of this field. After that, we will introduce the concept of metacognition ability by examining and modeling the decision-making process of humans. Then, we will describe the use of metacognition in crowdsourcing systems. The proposed model of the crowdsourcing system, considering the first and second types of decision-making, is introduced in Section \ref{Decision making model}. Through mathematical and experimental analysis, we will examine our proposed system. According to our findings, the crowdsourcing system error can be decreased by asking people how confident they are and khowing their metacognition ability. Furthermore, this approach allows us to avoid repetitive pre-tests and recruit users based on their metacognitive abilities, which are assessed at the beginning of their enrollment.

\section{Background} \label{literature review} 
\subsection{Crowdsourcing Systems}
Crowdsourcing is a method that relies on collective wisdom to answer questions. Usually, this method is used for questions whose answer is unavailable or there is not enough confidence in the available response. To obtain the answer, the desired question is asked from several people, and based on the crowd's response, the final answer is extracted. There are different types of crowdsourcing systems, some of which are listed below:
\begin{itemize}
	\item In some crowdsourcing systems, only one person provides a solution for each problem; For example, in \href{https://www.innocentive.com/}{InnoCentive} and \href{https://www.kaggle.com/}{kaggle}, companies describe their issues and questions. The people in this network offer a solution after studying the problem. At the end, the crowd answer will be extracted by an answer aggregation algorithm, which will be explained later. The users whose answers are the same as the crowd answer will be rewarded.
	
	\item Another group of crowdsourcing systems is designed to use data sent from mobile phones or computers \cite{sheng2013sensing}. In this type of system, there is no need for people to answer questions. One example of this category is traffic estimation by Google Map. This software receives the speed of cars from the sensors of smartphones and using this information and data obtained in the previous days and hours with the same method, it estimates the traffic and announces the fastest route to destination. In this type of systems, people are not given cash rewards, but they can use the data obtained by these softwares (for example, routing according to traffic). Another example of this area is finding a network map and classifying points in terms of antenna quality. For this purpose, a software is installed on the smartphone, and this software sends information about the network to a server in each region \cite{faggiani2014smartphone}.
	
	\item In another type of crowdsourcing system, a problem arises that is difficult to find an answer to or requires an expert. Since it is expensive to get help from an expert, crowdsourcing suggests that we leave this task to a large number of non-experts and then use their answer set. In the rest of this paper, a crowdsourcing system is a system that uses the method of assigning a task to a large number of people.
\end{itemize}
One of the most famous crowdsourcing system is \href{https://www.mturk.com/}{Amazon Mechanical Turk} (AMT). This site is a web marketplace that helps companies find people to solve problems that computers can't solve; For example, identifying and categorizing photos and videos and annotating them, converting images into documents and books \cite{liu2010crowdsourcing}, implementing the text of podcasts, etc. These tasks are called HITs. Each HIT is designed in such a way that it is not necessary to spend a lot of time doing them.

One of the uses of crowdsourcing is categorizing and annotating images and videos. Due to the large number of images available on the web, database, etc., this work is not possible manually because it is time consuming. On the other hand, for obtaining the best result, experts are needed and recruiting them is usually expensive. Sometimes techniques can be used that automatically annotate multimedia content using algorithms. The main disadvantage of these methods is that they only support limited vocabulary for annotation, and in complex cases, they lead to high errors \cite{ntalianis2014automatic}. For annotation complex cases, we can use an expert which is expensive. Therefore, one of the standard solutions is to use crowdsourcing in this field. For example, \cite{von2004labeling} has been able to label many images by designing a game. The paper predicts that if the game is played continuously by 5,000 people, all the photos on Google (425,000,000 photos in 2005) will be tagged within 31 days. In this game, participants are paid cash, and people play for fun only. Finally, a label that a certain number of people agree on is selected for each photo. It is also possible to classify images by quality using crowdsourcing \cite{ribeiro2011crowdsourcing} or improve their quality \cite{ghadiyaram2015massive}.

Another common application of these systems is to obtain a complete and accurate map inside the buildings \cite{shin2015participatory}. CrowdMap is a software designed for finding indoor maps which can help a variaty of jobs such as interior designers \cite{chen2015crowd}. This software asks the user to film the space inside the building. Then, a map inside the building is obtained using the people's geographic location and the video sent. Other uses of crowdsourcing include data collection for problem analysis. For example, \cite{mcduff2012crowdsourcing} shows different ads to people using crowdsourcing software. Then by analyzing their facial expressions and the questions , the system tries to provide a way to determine the impact of ads on people. Crowdsourcing systems have made it possible for different people to participate in business activities. Commercial companies can conduct market research, development, and testing of new products, pricing, etc., through these systems \cite{gatautis2014crowdsourcing}.

\subsubsection{Steps of a crowdsourcing system} 
In a crowdsourcing system, first, a task is defined. Then the system announces the task's information to all the crowdsourcing network's people. The task information is its topic and category, description, the time required to do it, etc. Then the users who are enthusiastic about doing this task announce their willingness. In some systems, users propose an amount they want to be paid in exchange for performing the desired task. But sometimes, the system calculates and announces the payment amount. Next, the system holds a pre-test. Based on predefined criteria (such as performance in the pre-test section), the system decides which users to recruit to contribute to the task. After this step, the task will be fully announced to the selected users. After completing the task, the system collects the answers, and the crowd's response is calculated using an answer aggregation method. Rewards will be paid to the people whose answers are recorded in the assigned time and whose answers are the same as the crowd's answer.

\textbf{User Recruitment} Selecting users is done in two main ways: online and offline \cite{zhang2014free}. In the offline method, the system waits to find the required number of users and then announces the task to everyone. For this mode, it is assumed that there are a number of simultaneous users in the system. One of the crucial things in this selection is predicting the location of people so that we can estimate how many people will be present in the situation where we want the task to be done and adjust the filters for selecting people \cite{liu2017prediction, pu2016crowdlet}.\\
Since offline recruiting is not easy, the online method is usually used. In this method, any user can announce his willingness in a period of time, and the system should decide whether to recruit the user or not.

\textbf{Answer Aggregation Algorithms.} In order to extract the answer from the crowd, different algorithms can be used. The two main categories of them are mentioned below:
\begin{itemize}
	\item Majority voting (MV): In this method, an answer is announced as the final answer that the majority of the crowd has agreed on \cite{liu2012cdas}. This method (as well as weighted majority voting) is the most common in these systems. That is because this method is easy to implement and the calculation overhead is much less compared to the other methods.
	\item Algorithms based on machine learning and pattern recognition: these algorithms are one of the main answer aggregation algorithms in the field of annotation. For example, \cite{nguyen2019explainable} found the final answer through clustering. Also, in \cite{yadav2011smsassassin}, it detects spam text messages using SVM and Bayesian filters. Naive Bayes \cite{kamar2012combining} or deep Bayesian \cite{yang2018leveraging} can be used to find suitable photo labels.
\end{itemize}

\textbf{Incentive mechanisms.} Users' rewards can be cash or non-cash. Some crowdsourcing systems are designed in such a way that participation in them is enjoyable (such as systems developed in the form of a game) or ultimately provide information and services to the users. But considering that doing the task will be costly for the person (use of mobile phone charging, internet, time, etc.), in some systems, a cash bonus is paid to the users.\\
There are different solutions for determining cash rewards. as an example, paying the winner is one of the incentive methods. This method puts people into a competitive phase and prevents them from cooperating with each other. Although the non-cooperation of users causes different responses and comments to be registered, it may reduce the number of users willing to do the task because they are no different in this second and last place structure. Therefore, except for people who are very interested in competition, other people will not have the motivation to participate \cite{abernethy2011collaborative}.\\
Another method is to pay a fixed amount to all the users \cite{nguyen2019explainable, gadiraju2017using}, which reduces the number of calculations to find the payment amount and gives everyone an incentive to participate in the task. Also, this method does not incentivize people to cheat for more money.\\

Auction is also used in these systems. In this case, the user submits his bid. Based on the bid price and budget, the system decides which users to use \cite{zhang2014free, zhao2014crowdsource}. Also, payment based on the required time and the difficulty of the question can be utilized \cite{bi2016optimal}.

\subsection{Meta-cognitive ability}
Decision-making requires gathering, organizing, synthesizing, and evaluating information. Due to each person's difference in understanding and intuition of the information needed to make a decision, people will not make the same decision in the same situations. Decision-making in people can be divided into two types: the first and the second type. The first type of decision-making refers to the individual's choice between the available options, while the second type refers to the degree of confidence that the decision has been made correctly \cite{mazor2022paradoxical}. The level of a person's ability to correctly estimate his level of confidence is called metacognition.\\

Metacognition refers to higher-order thinking that deals with the ability to influence and monitor one's cognitive processes, such as perception, memory, and decision-making. Metacognition usually deals with determining whether a person's certainty of answering can be a predictor of the correctness of their decision \cite{fleming2017hmeta}. Metacognition is more simply defined as "thinking about thinking." Metacognitive strategies are sequential processes that a person uses to control cognitive activities and ensure the achievement of a cognitive goal (for example, understanding a text). These processes help regulate and monitor learning and include planning and monitoring cognitive activities as well as examining the results of those activities. For example, after reading the paragraph in the text, each person can ask themselves about the concepts discussed in the paragraph. Their cognitive goal was to understand the text. Self-questioning is a metacognitive monitoring strategy \cite{livingston2003metacognition}.\\

As mentioned, metacognition plays a vital role in learning. People are often aware of their mistakes and report a level of confidence in their choices that is related to their performance. These metacognitive assessments of decision quality are essential for guiding behavior and subsequent decisions, especially when external feedback is absent \cite{fleming2017self}.\\

Suppose we consider a simple two-choice decision-making task. In that case, a person with high metacognition is assigned to one who declares his correct answers with high confidence and his incorrect answers with low confidence. In the presented models for decision-making, it is assumed that a decision variable is produced in the brain for each question. This variable is compared to the individual's decision threshold, and being greater or lower determines which option they choose. The difference between the decision threshold and decision variable will show one's level of confidence \cite{maniscalco2012signal}. If a person is asked to declare their confidence in a leveled and discrete manner, they will report this value according to the thresholds specified in their mind to declare confidence levels. The parameters involved in the production of the decision variable, the decision threshold, and the choice of confidence level are different from person to person.\\

There are various methods to study the first and the second type decision-makings. The most common method is using Signal Detection Theory (SDT) and using Receiver Operating Characteristic (ROC) \cite{maniscalco2014signal}, which are explained as follows. To understand how we calculate ROC first we should understand how people make decisions. The answer to any question that a person submits, based on the correct answer to the question, can have four states which are shown in Table \ref{tab:type1 decision}.

\begin{table}
	\begin{center}
		\caption{The possible modes of the first type of person's answer to two-choice questions}
		\label{tab:type1 decision}
		\begin{tabular}{lc|ll|}
			\cline{3-4}
			\multicolumn{2}{l|}{\multirow{2}{*}{}} & \multicolumn{2}{c|}{\textbf{Person's  response}} \\ \cline{3-4} 
			\multicolumn{2}{l|}{} & \multicolumn{1}{c|}{$a_1$} & \multicolumn{1}{c|}{$a_2$} \\ \hline
			\multicolumn{1}{|c|}{\multirow{2}{*}{\textbf{Correct   answer}}} & $a_1$ & \multicolumn{1}{l|}{Correct   rejection} & False alarm \\ \cline{2-4} 
			\multicolumn{1}{|c|}{} & $a_2$ & \multicolumn{1}{l|}{Miss} & Hit \\ \hline
		\end{tabular}
	\end{center}
\end{table}

Hit rate (HR) and false alarm rate (FAR) are used to examine the individual's behavior to answer the questions. If we draw a graph whose horizontal axis is FAR, its vertical axis is HR, and its points are the pairs (FAR, HR) and fit it, the obtained curve will be the first type ROC. If the person responds completely randomly, their ROC curve will be the line FAR=HR. Therefore, the area above this line shows a person's ability and expertise. The area under the first type ROC diagram (AUROC1) is directly related to performance (the rate of answering questions correctly).\\

To analyze Type II decision, assuming that there are only two levels of confidence (high and low), we would have four states for Type II decision-making,  shown in Table \ref{tab:type2 decision}.

\begin{table}
	\begin{center}
		\caption{The possible modes of the second type of person's answer to two-choice questions}
		\label{tab:type2 decision}
		\begin{tabular}{lc|ll|}
			\cline{3-4}
			\multicolumn{2}{l|}{\multirow{2}{*}{}} & \multicolumn{2}{c|}{\textbf{Person's   confidence}} \\ \cline{3-4} 
			\multicolumn{2}{l|}{} & \multicolumn{1}{c|}{High} & \multicolumn{1}{c|}{Low} \\ \hline
			\multicolumn{1}{|c|}{\multirow{2}{*}{\textbf{Answer}}} & Correct & \multicolumn{1}{l|}{Hit 2} & Miss 2 \\ \cline{2-4} 
			\multicolumn{1}{|c|}{} & Wrong & \multicolumn{1}{l|}{False alarm 2} & Correct rejection 2 \\ \hline
		\end{tabular}
	\end{center}
\end{table}

Similarly, we use HR 2 and FAR 2 to draw the second type ROC. The area below this diagram (AUROC2) represents a person's metacognition level. This method can be used for more confidence levels.

\subsection{Metacognition in Crowdsourcing Systems}
If we ask people for their confidence when asking crowdsourcing questions, we can calculate their metacognition and performance. Their confidence can also be used to aggregate votes. Existing examples of crowdsourcing systems do not typically use metacognition, and there are only a few examples in this field. One of the examples close to these systems using metacognition is a crowdsourcing system using self-assessment introduced in \cite{gadiraju2017using}.\\

It is stated in \cite{gadiraju2017using} that self-assessment can help attention and accuracy in answering questions in users. In the model presented by this article, after running a pre-test, it asks people what percentage of questions they have answered correctly according to their opinion. In addition to high efficiency, people will be recruited based on a relatively good performance estimation.\\

The Dunning-Kruger effect is a cognitive bias in which people assume their competency is low when they have high performance and vice versa. This effect and its influence on crowdsourcing systems are studied in \cite{gadiraju2017using}. This paper proposes a mathematical model for performance and confidence based on the DK effect. By assuming MV, and WMV aggregation algorithms, they have studied the impact of confidence reporting in crowdsourcing systems. They have concluded that WMV causes more errors than MV. According to our experiment, this finding is invalid, and the proposed model in this paper cannot be used in all cases.\\

In the next section, we will introduce and analyze our proposed crowdsourcing system. In this system, we will answer whether asking about the confidence level will improve the system's performance. To check this point, we will compare one of the common crowdsourcing systems with our proposed system.

\section{Decision-Making Model}\label{Decision making model}
As we mentioned in Section \ref{literature review}, there are two types of decisions: Type I and Type II. A Type I decision is a choice between the possible options. Type II decision represents the confidence level of the decision. In this section, we will discuss the mathematical model for these types based on what \cite{maniscalco2012signal} has presented. For the sake of simplicity, we will model the decision when humans face two-choice questions.

\subsection{Type I Decision}
\begin{figure}
	\begin{center}
		\includegraphics[width=0.57\textwidth]{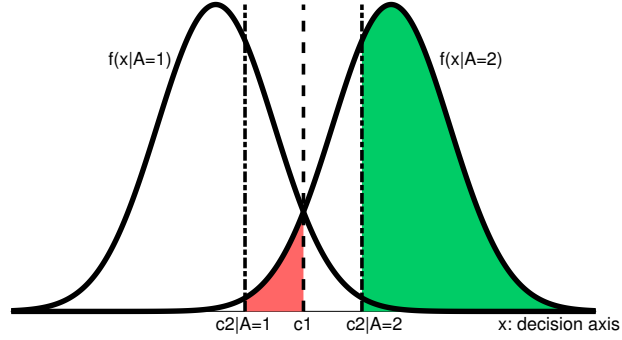}
	\end{center}
	\caption{Decision-making model. A normal random variable, $x$, is generated for each question. Type I decision is made based on the value of $x$ and the decision-making threshold, $c1$. The same happens with Type II decision, confidence, due to $x$ and Type II decision-making threshold, $c2|A$.
		\\The green area shows the probability of answering correctly if the answer is option two and the confidence is reported as high. Also, the red area is equal to the probability of being wrong, with the true answer being option 2, reporting low confidence.}
	\label{fig:decision_model}
\end{figure}

When a question is presented, a decision variable is generated corresponding to the evidence in favor of each option, $A=1$, and $A=2$. You will decide among the options based on the value of this variable and your decision-making threshold. A Gaussian model often models this random variable \ref{fig:decision_model}. The distribution's mean and variance vary from person to person depending on their performance and bias. The distribution of the decision variable, $x$, is as follows:
\begin{equation}\label{normal dist. x}
	x|A=i \sim \mathcal{N}(\mu_i, \sigma_i), \qquad i\in \{1,2\}
\end{equation}
Imagine you want to answer a question where the correct answer is $A = 1$. In your mind, a random variable that follows a Gaussian model is generated, $\mathcal{N}(\mu_1, \sigma_1)$. If the generated variable is lower than your Type I decision-making threshold, $c1$, you will choose $A=1$, otherwise $A=2$.

With the help of this model, the probability of answering correctly for users can be calculated, also known as performance. The $u$th user's performance, shown by $Perf_u$, is as follows:

\begin{align}
	Perf_u = & \Pr(A=1) \times \Pr(x\leq c1|A=1) \nonumber\\
	& + \Pr(A=2) \times \Pr(x> c1|A=2).  
\end{align}

Due to \eqref{normal dist. x}, both $\Pr(x\leq c1|A=1)$ and $\Pr(x> c2|A=2)$ can be written as $\erf(x) = \frac{1}{\sqrt{2\pi}}\int_{x}^{\infty}\exp(-\frac{u^2}{2}) du$. For defining this function, assume $Y$ is a Gaussian random variable with mean $\mu$ and variance $\sigma^2$, then $V=\frac {Y-\mu}{\sigma}$ is standard normal and
$$\Pr(Y>y) = Pr(V>v) = \erf(v) = \frac{1}{\sqrt{2\pi}}\int_{v}^{\infty}\exp(-\frac{u^2}{2}) du$$

\noindent Therefor, The $Perf_u$ is calculated as

\begin{align} \label{Perf_u}
	Perf_u  = &\Pr(A=1) \times [1-\erf(\frac{c1-\mu_1}{\sigma_1})] \nonumber \\
	& + \Pr(A=2) \times \erf(\frac{c1-\mu_2}{\sigma_2}).
\end{align}

\subsection{Type II Decision}
After choosing your answer to the question, you will decide how confident you are about the correctness of your decision; This is called Type II decision. This decision is similar to Type I, according to the decision variable, $x$, and Type II decision-making threshold. By assuming $c_{max}$ is the maximum level of confidence that can be reported, this threshold is $c2=j|A=i \quad i\in\{1,2\},~ j\in\{1,...,c_{max}\}$. 

Assuming we have two confidence levels (low and high), the levels are divided by one threshold. In the previous example, if your choice was $A=1$ and $x$ is greater than $c2|A=1$, your confidence is low; otherwise high. An example is shown in figure \ref{fig:decision_model}.\\

The meta-cognitive ability that was discussed in section \ref{Introduction} is how good the user's confidence ratings discriminate between their own correct and incorrect stimulus classifications \cite{maniscalco2012signal}. Several methods measures meta-cognition ability, but in this study, we use the area under Type II ROC curve, AUROC2 \cite{fleming2014measure}. The y-axis of this curve is Pr( confidence | correct) and the x-axis is $\Pr(\text{confidence} | \text{incorrect})$. Due to its definition, the greater AUROC2 means the user is more meta-cognitive.\\

All the parameters defined in this section differ among users and can vary over time. They all can be obtained by fitting this model to their decision-making behavior. In section \ref{Method_section}, we will use this model to make decisions in a crowdsourcing system.

\section{Method} \label{Method_section}
In any crowdsourcing system, after broadcasting the task information and collecting volunteers' requests, who desire to do the task, the system selects the most proper subset of volunteers. This selection is either on the features or skills each person claims or according to the pre-test results. The selected volunteers, also known as users, are recruited to participate in the test phase. In this phase, the main questions are presented to the users, and their answers are collected. The system extracts the crowd's answer based on an aggregation method and announces it as the correct answer. Users receive incentives at the end of the task due to the system's payment policy. In this paper, we assume incentives for all the users are the same. In this section, we first discuss about a conventional crowdsourcing system, \textbf{Re}sponse \textbf{Ba}sed \textbf{C}rowdsourcing \textbf{S}ystem (ReBaCS), and then our proposed system,\textbf{Co}nfidence-\textbf{Ba}sed \textbf{C}rowd-\textbf{S}ourcing system (CoBaCS) will be introduced.\\

\subsection{Response Based CrowdSourcing System (ReBaCS)}
As mentioned earlier, \textbf{Re}sponse \textbf{Ba}sed \textbf{C}rowdsourcing \textbf{S}ystem, abbreviated as \textbf{ReBaCS}, constitute a prevalent and conventional type wherein user responses form the basis for answer aggregation through the Majority Voting (MV) algorithm. Each ReBaCS undergoes two distinct phases: a pre-test phase evaluating volunteer capabilities, and a subsequent test phase where main questions are presented to recruited users. During the pre-test, users submit their answers while the system possesses knowledge of correct responses. User selection is contingent on performance, assessed by the accuracy of their answers. Proper user selection necessitates contextual alignment between pre-test questions and the main crowdsourcing task. Recruited users, meeting a minimum performance threshold in the test phase, proceed to answer main crowdsourcing questions. The crowd's answer to each question is determined by achieving consensus among more than half of the crowd. Compensation is provided to users upon task completion.

\subsubsection{Error Calculation}

For the sake of simplicity, we assume that all the questions in both the pre-test and test phases are two-choice. The answer of the $n$th user to question $q$ can be written as:
\begin{equation}
	r_{n,q} = \begin{cases}
		0 & \text{if the $n$th user has chosen the first option}\\
		1 & \text{if the $n$th user has chosen the second option}
	\end{cases}
\end{equation}

By assuming $N_u$ users have passed the pre-requisite test and are presented in the system, the crowd's answer for question $q$, $R_{\text{MV}_q}$, is driven by MV as
\begin{equation}
	R_{\text{MV}_q} = \begin{cases}
		0, & \sum_{n=1}^{N} r_{n,q} <\frac{N_u}{2},\\
		1, & \text{o.w},
	\end{cases}
\end{equation}

If we denote the correct answer of the $q$th question by $A_q$, the system's performance is the proportion of the questions in which their actual response, $A_q$, matches the crowd response, $R_{\text{MV}_q}$. If we assume $N_Q$ questions are available in the system, the system's accuracy is defined as the percentage of the correct answers as 
\begin{align}
	acc_{\text{ReBaCS}} &= \frac{\sum_{q=1}^{N_Q}\delta(R_{\text{MV}_q},A_q)}{N_Q}, \\
	\delta(x,y) &= \begin{cases}
		1,& \text{if } x=y,\\
		0, & \text{o.w.},
	\end{cases} \nonumber
\end{align}

By the above definition, the error of the system will be $err_{\text{ReBaCS}} = 1-acc_{\text{ReBaCS}}$. With the help of the total probability theorem, the expectation of error can be rewritten as follows:
\begin{align}
	\hspace{-8mm} \E_{N_Q}[err_{\text{ReBaCS}}] = &\Pr(A\neq R_{\text{MV}}) \\
	= &\Pr(A=0) \times \Pr(R_{\text{MV}}=1|A=0) \nonumber \\
	&+ \Pr(A=1) \times \Pr(R_{\text{MV}}=0|A=1) \nonumber \\
	= &\Pr(A=0) \times \Pr\left(\sum_{n=1}^{N} r_{n} \geq \frac{N_u}{2}|A=0\right) \nonumber \\
	&+\Pr(A=1) \times \Pr\left(\sum_{n=1}^{N} r_{n} < \frac{N_u}{2}|A=1\right). \nonumber
\end{align}

In these equations, $A$ shows the actual answer to a question, and $R_{\text{MV}}$ refers to the crowds' response based on the MV algorithm. According to the decision-making model described in Section \ref{Decision making model}, each user decides between the first and the second option in every question based on a Bernoulli random variable as:

\begin{align}\label{B random var}
	\hspace{-7mm} \Pr(r_n=r | A=i) = &\begin{cases}
		\erf\left(\frac{c_n-\mu_{i,n}}{\sigma_{i,n}}\right) & r=1 \\
		1-\erf\left(\frac{c_n-\mu_{i,n}}{\sigma_{i,n}}\right) & r=0
	\end{cases},  \ \ i \in \{0,1\}
\end{align}

\noindent where $\erf(x)$ is $\frac{1}{\sqrt{2\pi}}\int_{x}^{\infty}\exp(-\frac{u^2}{2}) du$, $c_{n}$ , $ \mu_{i,n}$ and $\sigma_{i,n}$ are the decision making parameters for the $n$th user, respectively. Each user's decision-making characteristics are encoded in how far the $\mu_{i,n}$s are from $c_n$ and the difference between $\mu_{i,n}$s. Therefore, without loss of generality, we can assume $c_n=0$ and $\sigma_{0,n}=\sigma_{1,n}=1$ for all users and let the value of $\mu_{i,n}$ show their decision making policies. By these assumptions, the equation in \eqref{B random var} becomes less complex:
\begin{align}
	\Pr(r_n=r | A=i) = &\begin{cases}
		\erf(\mu_{i,n}) & r=1 \\
		1-\erf(\mu_{i,n}) & r=0
	\end{cases}
\end{align}

The summation of several independent Bernoulli random variables with the same distribution parameters is a Binomial random variable. Therefore, by considering that all $\mu_{i,n}$ are the same and equal to $\E[\mu_{i,n}]=\mu_{i}$, the random variables defined as $\gamma_{i}=\sum_{n=1}^{N_u} r_n |\big(A=i\big), ~i\in\{0,1\}$ are binomial random variables. If we suppose $N_Q$ is large enough, $\gamma_i$ follows a normal distribution as

\begin{align}\label{N random var}
	\gamma_{i} &\sim \mathcal{N}(m_{i,r}, s_{i,r}) \\
	&\qquad m_{i,r}  = N_u\times \erf(-\mu_i) \nonumber\\
	&\qquad s_{i,r}  =\sqrt{N_u\times \erf(-\mu_i)\times\big(1-\erf(-\mu_i)\big)}\nonumber
\end{align}

\noindent With the equation in \eqref{N random var}, $\E[E_{ReBaCS}]$ is calculated as:
\begin{align}\label{E ReBaCS}
	\E[err_{\text{ReBaCS}}]
	=& \Pr(A=0) \times \Pr\left(\frac{N_u}{2} \leq \gamma_0\leq N_u\right) \nonumber \\
	& + \Pr(A=1) \times \Pr\left(0 \leq \gamma_1 <\frac{N_u}{2}\right) \nonumber \\
	= &\Pr(A=0) \times \Big(\erf\left(\frac{\frac{N_u}{2}-m_{0,r}}{s_{0,r}}\right) -\erf\left(\frac{N_u-m_{0,r}}{s_{0,r}}\right)\Big) \nonumber \\
	& +\Pr(A=1) \times \Big(\erf\left(\frac{-m_{1,r}}{s_{1,r}}\right) -\erf\left(\frac{\frac{N_u}{2}-m_{1,r}}{s_{1,r}}\right)\Big) \nonumber \\
\end{align}

In conclusion, by knowing the expectation of all users decision variables and the number of users, expectation of ReBaCS error can be determined.

\bigskip
\subsection{CoBaCS}
As we discussed earlier, sometimes users are not 100\% sure about the correctness of their answers, so they choose the option they are more confident about. By knowing the meta-cognitive ability of each user, we can estimate how accurately their confidence leads them. In other words, we can answer this question: "Is the user's confidence a proper estimator of his/her accuracy?". If we select volunteers based on their meta-cognitive ability and, in every question, ask them to submit their confidence as well as their answers, we can improve system performance. So, in the \textbf{Co}nfidence-\textbf{Ba}sed \textbf{C}rowd-\textbf{S}ourcing system, \textbf{CoBaCS}, first we show the volunteers pre-test questions. As we mentioned earlier, both their response and confidence are recorded. At the end of this phase, users are selected due to their both performance and meta-cognition ability. For measuring performance, the context and the complexity of the pre-test and test phase questions must be the same. Still, according to  \cite{veenman1997generality,mazancieux2018retrospective} meta-cognitive ability is similar across domains. So, the developed tasks in the literature can be applied, and there is no need to re-design them.

In the test phase, the same as the pre-test, users report their confidence in being correct in each question. After collecting users' answers and confidence, the system states the crowd answer by Weighted Majority Voting (WMV) algorithm. This algorithm works similarly to MV with one difference; answers are multiplied with their corresponding confidence. Using WMV, a user with low confidence has a lower contribution to the crowd's answer.

\subsubsection{Error Calculation}
In CoBaCS, we assume $r_{n,q}$ will be either 1 or -1 as

\begin{equation}
	r_{n,q} = \begin{cases}
		-1 & \text{if the $n$th user has chosen the first option}\\
		1 & \text{if the $n$th user has chosen the second option}
	\end{cases}
\end{equation}
Users are asked to report their confidence as discrete levels from 1 (i.e., lowest confidence) to $c_{max}$ (i.e., highest confidence). We denote the reported confidence of user $n$ to the $q$th question by $c_{n,q}$. Then, we have
\begin{equation}
	R_{\text{WMV}_q} = \begin{cases}
		-1 & \text{if} \ \sum_{n=1}^{N} r_{n,q}\times c_{n,q} <0\\
		1 & \text{o.w.}
	\end{cases}
\end{equation}
Similar to the error calculation of ReBaCS, CoBaCS performance is the fraction of questions that the crowds' consensus leads the system to the correct answer. The expectation of error in this system is as follows:
\begin{align}
	\E_{N_Q}[err_{\text{CoBaCS}}] = &\Pr(A=-1) \times \Pr(R_{\text{WMV}}=1|A=-1) \nonumber \\
	&+ \Pr(A=1) \times \Pr(R_{\text{WMV}}=-1|A=1)\nonumber\\
	= &\Pr(A=-1) \times \Pr\left( \sum_{n=1}^{N} r_{n}\times c_{n} \geq0 \left|A=-1\right.\right)\nonumber \\
	& + \Pr(A=1)\times \Pr\left( \sum_{n=1}^{N} r_{n}\times c_{n} <0\left|A=1\right.\right)\nonumber
\end{align}

Due to the independency of $r_{n}\times c_{n}$ across $n\in\{1,... , N_u\}$ and their limited variance, the sum of these variables can be approximated with a normal random distribution as

\begin{align}
	\lambda_i & = \sum_{n=1}^{N_u}\big(r_{n}\times c_{n}\big)|\big(A=i\big), \qquad i\in\{-1,1\} \\
	&\qquad\lambda_i  \sim \mathcal{N}(m_{i,c},s_{i,c})\nonumber\\
	&\qquad m_{i,c}   = N_u\times \E[r\times c|A=i] \nonumber \\
	&\qquad s_{i,c}    = \sqrt{N_u\times \Var[r\times c|A=i]}\nonumber
\end{align}
For determining the error of CoBaCS, the confidence level threshold of users is needed as well as their first-order decision variables and the number of users. By having these parameters, the expected error of CoBaCS can be written as follows.

\begin{align}\label{E_CoBaCS}
	\E[err_{\text{CoBaCS}}] = &\Pr(A=-1) \times \Pr(\lambda_1>= 0) \\
	& + \Pr(A=1) \times \Pr(\lambda_2<0)\nonumber\\
	= & \Pr(A=-1) \times \erf(\frac{-m_{1,c}}{s_{1,c}})\nonumber\\
	& + \Pr(A=1) \times (1-\erf(\frac{-m_{2,c}}{s_{2,c}}))\nonumber
\end{align}

\subsection{Evaluations Environment}

In this section, we evaluate the performance of these systems and compare them in terms of error.To accomplish this, we simulate the recruited population of these cases.
For computing error of ReBaCS, we consider that the recruited population is made of users with various performances. We choose the performance of each user randomly. With having users' performances, the parameters of Type I decision making described in Section \ref{Decision making model} can be computed ($\mu_{i,n} \quad i\in\{0,1\}, n\in\{1,..,N\}$). Then, the expected parameters of decision-making will be obtained ($\mu_{i,r}\quad i\in\{0,1\}$).

We assume each user has either low, medium or the high meta-cognitive ability for meta-cognition ability without loss of generality. A user with low meta-cognitive ability is one who rates his or her confidence low in the time of being correct and high when the answer is wrong (AUROC2 $\approx0$). The high meta-cognitive user acts the contrary: low confidence in wrong answers and high confidence in correct answers (AUROC2 $\approx1$). The medium meta-cognitive user is one who consistently rates confidence at the medium level (AUROC2 $=0.5$). The users' meta-cognition ability is chosen randomly. 

\begin{figure}
	\centering
	\includegraphics[width=0.6\textwidth]{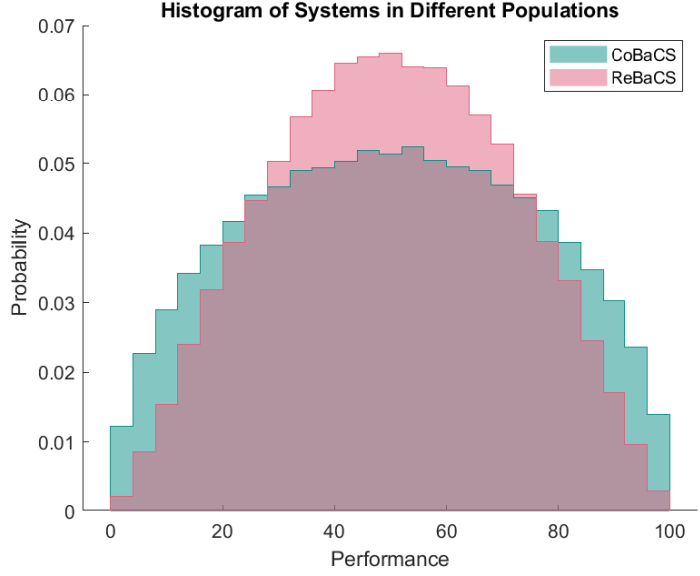}
	\caption{Performance of ReBaCS and CoBaCS in various populations: without filtering users at the beginning of the task, CoBaCS has no superiority over ReBaCS.}
	\label{fig:simulation performance}
\end{figure}

We assumed that 200 questions and 100 users were participating in both ReBaCS and CoBaCS systems. We have chosen these numbers to have enough questions to calculate the system's error correctly. Also, We aimed to have a reasonable number of users remain in the system after filtering. Considering 100,000 samples of the population with different performances and meta-cognitive ability, Figure \ref{fig:simulation performance} is obtained. As you can see in this figure, without having any filters on performance and meta-cognitive ability, based on performance, CoBaCS is in 52.38 \% of viable populations better the ReBaCS. Although in some populations, CoBaCS outperforms ReBaCS, in general, without any pre-testing and filtering, there is no significant difference between ReBaCS and CoBaCS systems.

\begin{figure*}
	\centering
	\begin{subfigure}[b]{0.6\textwidth}
		\centering
		\includegraphics[width=\textwidth]{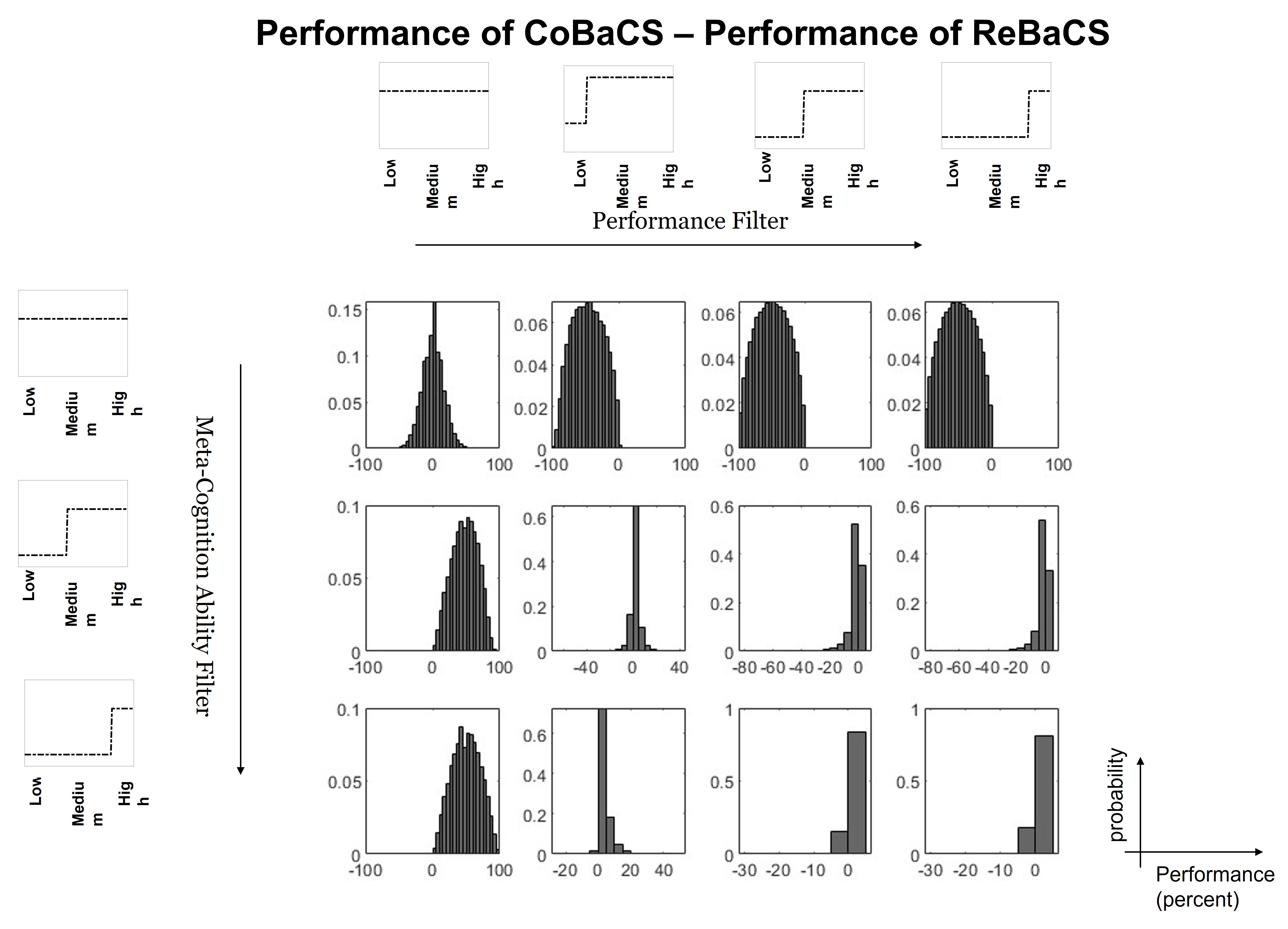}
		\caption{}
		\label{fig:difference plots}
	\end{subfigure}
	\hfill
	\begin{subfigure}[b]{0.6\textwidth}
		\centering
		\includegraphics[width=\textwidth]{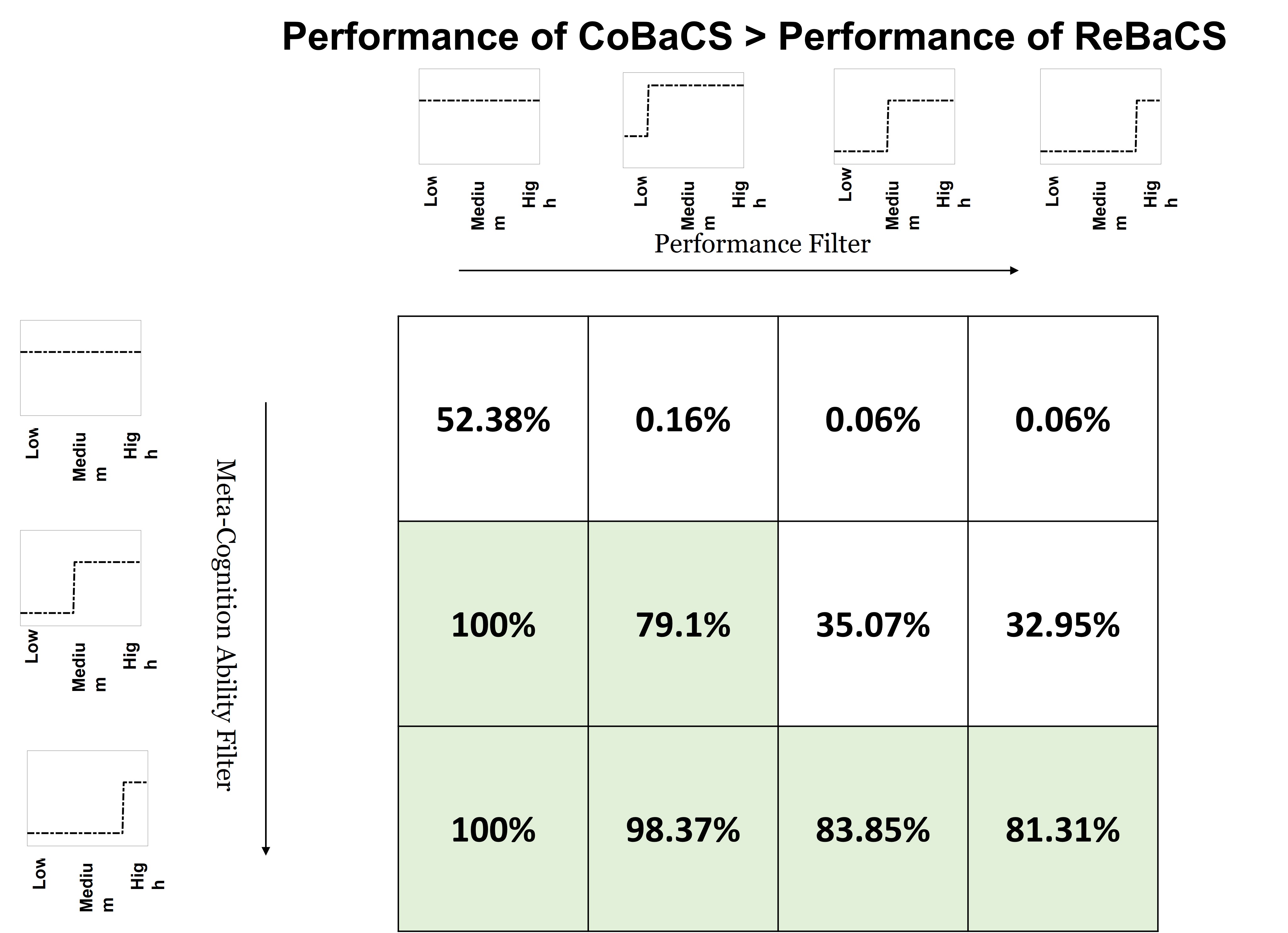}
		\caption{}
		\label{fig:difference table}
	\end{subfigure}
	\caption{ReBaCS and CoBaCS exhibit divergent performance patterns based on user populations. CoBaCS excels when low meta-cognitive ability users are absent, while ReBaCS outperforms in the presence of expert users. The study employs histograms and percentages to illustrate these performance dynamics across various filters.}
	\label{fig:simulation performance difference}
\end{figure*}

In Figure \ref{fig:simulation performance difference}, we have demonstrated the effect of different filters in the pre-test phase on systems' performance. Users are filtered according to their performance in the horizontal direction and their meta-cognition ability in the vertical direction. According to this figure, in the populations where the experts (users with high performance) are present, it is wiser to use ReBaCS if we have no hard meta-cognition filter. But, when the users are not experts and have low performance, we can obtain better performance by filtering very low metacognitive users using CoBaCS. In the first column of Figure \ref{fig:simulation performance difference}, there are no criteria for users' performance, and with existing users, even with performance near 10\%, the CoBaCS is much better when we filter very low metacognitive users.

In conclusion, according to the simulations' results, when the CoBaCS is used, and the population does not include experts, it is important to filter people with low meta-cognition ability. It is noteworthy to mention that for filtering users based on their performance, we should take pre-tests in the context of the main task, which is challenging. On the other side, the metacognition ability is not task dependent \cite{veenman1997generality,mazancieux2018retrospective}; we can compute users' metacognition first, and then assign different task to the ones with no low metacognition ability. Also, in difficult tasks, finding users who are expert is hard and time-consuming. But, if they are available with hard metacognition filters, CoBaCS has a lower error. Therefore, CoBaCS can increase performance and reduce the cost in the time of difficulty. In Section \ref{Experimental}, we will examine our proposed system with a real-world task.

\section{Experimental Results and Analysis}\label{Experimental}
\subsection{Tasks and Procedures}

We consider three tasks to study the role of meta-cognitive ability as a filter and the effect of confidence reporting in crowdsourcing systems. The first task is a tweet task in which the participants are asked to guess the gender of the Twitter account owner by seeing a few tweets of him/her. The tweet task is designed to test the performance of a confidence-based crowdsourcing system, CoBaCS. Before the tweet task, we ran a pre-tweet task designed to evaluate the gender bias in different topics to rank the questions in the tweet task based on their difficulty. The other task is a word recognition memory task which was used to measure the meta-cognitive ability of subjects \cite{baird2013medial, Sadeghi2017}. The pre-tweet task was an online test using Google Forms. The memory and tweet tasks were computerized and programmed in MATLAB (Mathworks), using COGENT 2000 toolbox. \footnote{\href{http://www.vislab.ucl.ac.uk/cogent.php}{http://www.vislab.ucl.ac.uk/cogent.php}} The datasets gathered and used during the current study will be available from the corresponding author on reasonable request.\\

\textbf{Subjects:} Eighty-six participants completed the pre-tweet task. To gather these participants, we posted an open invitation on social media, welcoming everyone to join the task. Since there were no prior selection criteria, the users are assumed to have been selected randomly, eliminating the need for additional action to ensure the study's validity. The participants included 86 individuals (42 women, 44 men) with a mean age of 24.24 (range = 17 to 37), who were then instructed to complete the other two tasks. No participants were excluded from this study. All participants received a fixed payment of 150,000 Iranian Rials, ensuring a consistent incentive. Additionally, participants were ranked based on their performance in the Memory task combined with twice their performance in the tweet task. The top three performers in this ranking received a bonus of 100,000 Tomans. Participants were initially informed only about the duration of the task and that it would be completed on a computer. Detailed instructions regarding the procedure were provided just before they began the test. Informed consent was obtained from all participants. The study was approved by the Research Ethics Committees of the Faculty of Psychology and Education, University of Tehran (approval ID: IR.UT.PSYEDU.REC.1403.025), and the experiments were conducted in accordance with relevant guidelines and regulations.\\

\begin{figure}
	\centering
	\includegraphics[width=0.8\textwidth]{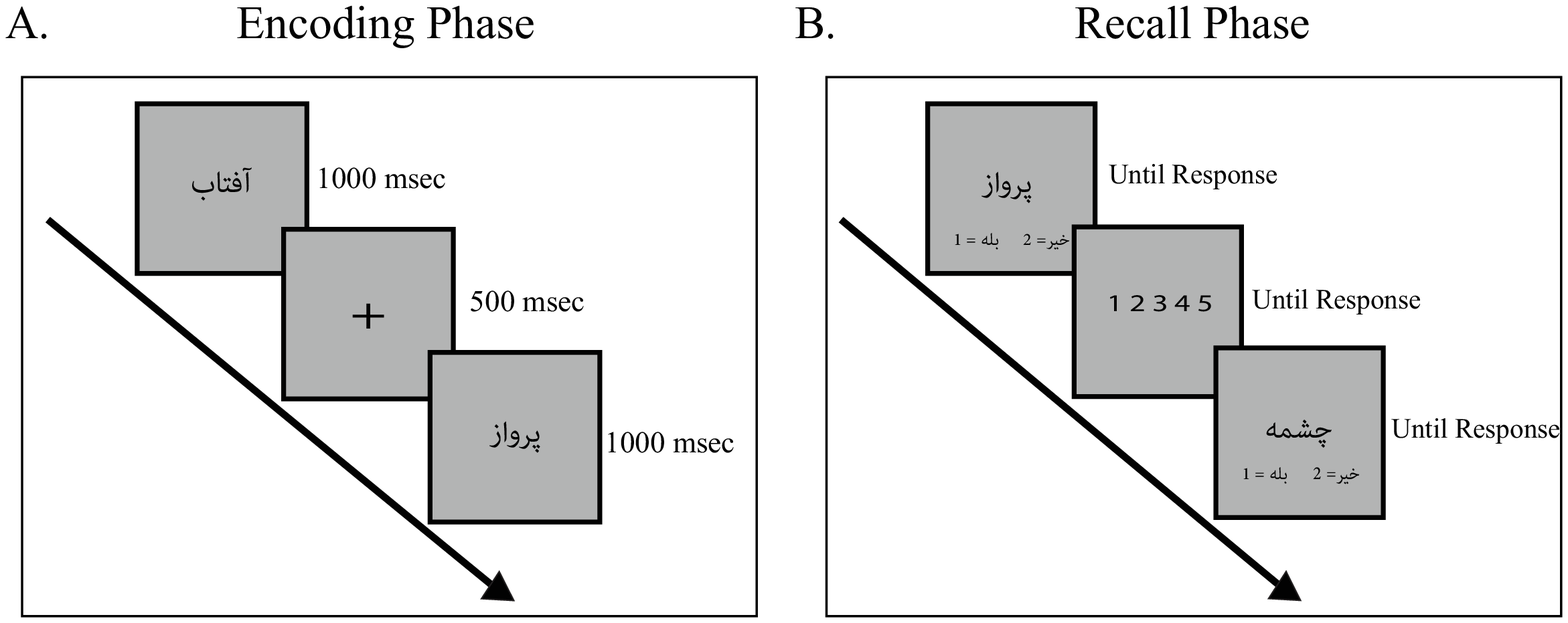}
	\caption{The Memory Task comprised two phases: an encoding phase where subjects memorized a random selection of words and a recall phase where subjects reported whether a presented word was seen in the encoding phase and rated their confidence in their response. Persian words, representing concepts like the sun rising, flight, and fountain, were used in the task.}
	\label{fig:mem_task}
\end{figure}

\textbf{Memory task:} The memory task was a Persian word recognition memory test based on \cite{baird2013medial} and adapted from \cite{Sadeghi2017}. The study had two phases: encoding and recall. During the encoding phase, 100 words from a set containing 200 words were presented on the screen in a random order, separately and sequentially. In the recall phase, immediately after the first phase, all 200 words were displayed sequentially. Subjects decided whether each word was shown in the encoding phase or not by pressing 1 (yes) or 2 (no) on the keypad using their left hand. Next, they reported their confidence in being correct on a scale of 1 (low confidence) to 5 (high confidence) using a number pad with their right hand. The procedure is shown in Figure \ref{fig:mem_task}. During the expemriment, Individuals did not receive any feedback on the correctness of their responses.\\

In the memory task, average memory performance of our subjects was 67.53\% (standard deviation  7.42\%, range 53\% -86.5\% ), comparable to the results reported in \cite{baird2013medial} (mean accuracy 71\%, standard deviation 9\%, range 57-91\%) and the results reported in \cite{Sadeghi2017} (mean accuracy 71\%, standard deviation 1\%, range 68 - 73 \%).

The average meta-cognitive ability of subjects was 0.62 (standard deviation 0.06, range 0.508- 0.784). Reported results on meta-cognitive accuracy were measured by area under ROC Type II curve, AUROC2 \cite{fleming2014measure, veenman1997generality, maniscalco2012signal}. \\

\textbf{Pre-tweet task.} The goal of this task was to investigate how difficult it is to guess the gender of the person behind a Twitter account based on the topics that have been discussed in their tweets. We designed an online questionnaire using Google Forms, including 17 titles. Sixty eight participants answered the questioner; they were instructed to report \emph{male} or \emph{female} for each of the contents. There was not any time limit to answer the questionnaire.

We define the gender of a tweet based on the gender that more than 60\% of the participants picked. A simple tweet is one in that the gender of the tweet is the same as the actual gender of its Twitter account. By the results of this task, we categorized selected tweets for the next task, the tweet task, into four groups based on the actual gender of the writer of tweets (female or male) and the difficulty of gender recognition according to the discussed topic (simple or complicated).

\textbf{Tweet task.} We collected a set of thousands of Twitter accounts that posted tweets in Persian. Then, one hundred accounts were selected, with more than one hundred followers and one hundred followings, and also had tweeted more than five hundred tweets. The last condition is to make sure that the accounts' owners are familiar with tweeting and that we have enough tweets from them to choose for our task. We omitted the accounts that had a massive number of tweets, but they have joined Twitter in recent months from our selections to avoid bots. Half of the accounts' owners were women, and the other half were men. We chose three of the tweets of these one hundred accounts sudo-random; thus, we had three hundred tweets in the Persian language. The selection of accounts and tweets was that every four categories of Twitter accounts (female/male and difficult/simple) occurred exactly 25 times (a quarter of the total number of accounts). An account was labeled "simple" if the majority of its selected tweets were deemed simple, according to the definition established in the pre-tweet task.

\begin{figure}
	\begin{center}
		\includegraphics[width=0.4\textwidth]{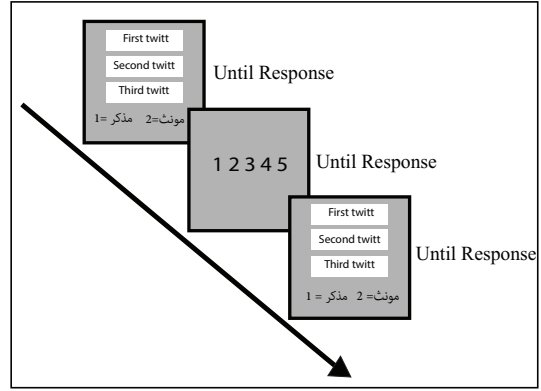}
	\end{center}
	\caption{Tweet Task. This Task had 100 questions, and each question contained three tweets, all from an account. The subjects were asked to guess the gender of the owner of the account. Then, they rated their confidence in being correct from 1 to 5.}
	\label{fig:twitt_task}
\end{figure}

At first, all the participants were asked to submit their gender and age. Participants of the tweet task saw the three tweets of each account on the same page on the monitor, and they decided whether the account belonged to a male or female by pressing the key "1" or "2". Then, they reported their confidence of correctness by pressing keys 1 (i.e., lowest confidence) to 5 (i.e., highest confidence) on the number pad (Figure \ref{fig:twitt_task}). There was no time limit to report the decision and confidence. At the end of the task, participants reported their estimation of the number of correct answers; then, they saw their actual performance. \\

In the tweet task, the average performance of our subjects was 61.79\% (standard deviation 4.56\%, range 49\%-72\% ). The mean of meta-cognitive ability was 0.58 (standard deviation 0.07, range 0.445- 0.777). In the literature, There is no similar task to our tweet task, so a comparison of results is impossible.\\

\subsection{Behavioral Results}

\textbf{Data Analysis:} We would like to study the influence of the performance filter and Metacognition filter on the error of two systems; ReBaCS and CoBaCS. Thus we needed two scores for each participant regarding performance and metacognitive ability. We saw that participants' performance in the first thirty trials of the tweet task was significantly correlated (Pearson correlation $= 0.59$, $p < 0.05 $) with their performance in the remaining questions, the last seventy, of the tweet task. Therefore, we used the performance in the first thirty questions (including 15 questions about men tweet applicants and 15 questions about women tweet applicants) of the tweet task as each participant's performance score. For the score of metacognitive ability, we used the AUROC2 for each participant in the memory task. It is noticeable that the first thirty trials of the tweet task were not enough to measure metacognitive ability in the tweet task to use it as the score of metacognitive ability; thus, we measured this ability in the memory task.\\

Then, we randomly sampled 60\% of participants, and we used four filters for performance ability ($0.4,0.5,0.6,0.7$) and three filters for metacognitive ability ($0.5,0.6,0.7$). Thus we applied twelve coupled filters on the 60 sampled participants. For example, for the coupled filter (performance 0.4, metacognition 0.5), we excluded the participants whose performances were lower than 0.4 and whose metacognitive abilities were lower than 0.5. To achieve the performance of the ReBaCS system, we saw the performance of the remaining people in the society after filtering in the last seventy questions in the tweet task according to the majority voting; society's answer to each question was the answer of the majority. Similarly, to achieve the performance of the CoBaCS system, we applied the WMV algorithm to measure the performance of the filtered society. The difference in society performance between the two MW/WMV algorithms and the 100\% performance showed the error of two crowdsourcing systems, ReBaCS and CoBaCS. In this way, we estimated the error of two crowdsourcing systems with twelve filters for a randomly sampled society with 60 participants. We repeated the above process 100 times, randomly sampled a new society with 60 percent, and measured the error of two systems after implementing the twelve filters. The results of this analysis is shown in Table \ref{tab:experiment outcome} and Table \ref{tab:experiment outcome2}. After obtaining the results, we ran a one-tailed T-test to check if the ReBaCS error was significantly greater than the CoBaCS error. The p-value for the first meta filter (meta>0.5) and performance greater than 40\% was 0.03. For the same meta condition and performance greater than 70\%, the p-value was 0.01. For the rest, this value was below $2e-5$. Therefore, for all conditions, the error of CoBaCs is significantly less than ReBaCS in this experiment. To test our hypothesis with the filter effect, we used a one-tailed T-test, and the result showed that regardless of the filter, the error of ReBaCS is significantly greater than the CoBaCS (p-value = $2e-100$).

\begin{table*}
	\begin{center}
		\caption{ReBaCS and CoBaCS Experiment Outcomex: Comparison of the Error of Systems.}
		\begin{tabular}{ccc|cccc|}
			\cline{4-7}
			\multicolumn{3}{c|}{\multirow{2}{*}{\textbf{Filter   Parameters}}} & \multicolumn{4}{c|}{\textbf{Min of Performance}} \\ \cline{4-7} 
			\multicolumn{3}{c|}{} & \multicolumn{1}{c|}{\textbf{40\%}} & \multicolumn{1}{c|}{\textbf{50\%}} & \multicolumn{1}{c|}{\textbf{60\%}} & \textbf{70\%} \\ \hline
			\multicolumn{1}{|c|}{\multirow{6}{*}{\textbf{Min of Meta}}} & \multicolumn{1}{c|}{\multirow{2}{*}{\textbf{0.5}}} & ReBaCS & \multicolumn{1}{c|}{38.5 $\pm$ 1.5} & \multicolumn{1}{c|}{32.5 $\pm$ 1.6} & \multicolumn{1}{c|}{32.2 $\pm$ 1.9} & 32.2 $\pm$ 2.6 \\ \cline{3-7} 
			\multicolumn{1}{|c|}{} & \multicolumn{1}{c|}{} & CoBaCS & \multicolumn{1}{c|}{32.0 $\pm$ 1.5} & \multicolumn{1}{c|}{31.8 $\pm$ 1.4} & \multicolumn{1}{c|}{31.5 $\pm$ 1.9} & 32.0 $\pm$ 2.8 \\ \cline{2-7} 
			\multicolumn{1}{|c|}{} & \multicolumn{1}{c|}{\multirow{2}{*}{\textbf{0.6}}} & ReBaCS & \multicolumn{1}{c|}{32.9 $\pm$ 2.1} & \multicolumn{1}{c|}{33.6 $\pm$ 2.2} & \multicolumn{1}{c|}{33.7 $\pm$ 2.2} & 34.9 $\pm$ 3.4 \\ \cline{3-7} 
			\multicolumn{1}{|c|}{} & \multicolumn{1}{c|}{} & CoBaCS & \multicolumn{1}{c|}{30.9 $\pm$ 2.1} & \multicolumn{1}{c|}{31.9 $\pm$ 2.1} & \multicolumn{1}{c|}{32.5 $\pm$ 2.3} & 33.9 $\pm$ 3.8 \\ \cline{2-7} 
			\multicolumn{1}{|c|}{} & \multicolumn{1}{c|}{\multirow{2}{*}{\textbf{0.7}}} & ReBaCS & \multicolumn{1}{c|}{37.1 $\pm$ 2.7} & \multicolumn{1}{c|}{37.6 $\pm$ 2.8} & \multicolumn{1}{c|}{41.5 $\pm$ 18.5} & - \\ \cline{3-7} 
			\multicolumn{1}{|c|}{} & \multicolumn{1}{c|}{} & CoBaCS & \multicolumn{1}{c|}{32.5 $\pm$ 3.0} & \multicolumn{1}{c|}{37.7 $\pm$ 3.5} & \multicolumn{1}{c|}{37.0 $\pm$ 20.3} & - \\ \hline
		\end{tabular} \label{tab:experiment outcome}
	\end{center}
\end{table*}

\begin{table}
	\begin{center}
		\caption{RebaCS and CoBaCS Experiment Outcomes: Mean of the number of subjects remaining after filtering}
		\begin{tabular}{ll|llll|}
			\cline{3-6}
			\multicolumn{2}{l|}{\multirow{2}{*}{\textbf{Filter Parameters}}} & \multicolumn{4}{l|}{\textbf{Min of Performance}} \\ \cline{3-6} 
			\multicolumn{2}{l|}{} & \multicolumn{1}{l|}{\textbf{40\%}} & \multicolumn{1}{l|}{\textbf{50\%}} & \multicolumn{1}{l|}{\textbf{60\%}} & \textbf{70\%} \\ \hline
			\multicolumn{1}{|l|}{\multirow{3}{*}{\textbf{Min of Meta}}} & \textbf{0.5} & \multicolumn{1}{l|}{52} & \multicolumn{1}{l|}{49.6} & \multicolumn{1}{l|}{36} & 8.4 \\ \cline{2-6} 
			\multicolumn{1}{|l|}{} & \textbf{0.6} & \multicolumn{1}{l|}{30.9} & \multicolumn{1}{l|}{29.8} & \multicolumn{1}{l|}{22.3} & 5.4 \\ \cline{2-6} 
			\multicolumn{1}{|l|}{} & \textbf{0.7} & \multicolumn{1}{l|}{4.4} & \multicolumn{1}{l|}{4.2} & \multicolumn{1}{l|}{1.7} & 0 \\ \hline
		\end{tabular}\label{tab:experiment outcome2}
	\end{center}
\end{table}

\textbf{Discussion.} As you can see in tables \ref{tab:experiment outcome} and \ref{tab:experiment outcome2}, the mean of error of RebaCS is greater than the other system in almost every case, Even in the first filter (meta>0.5, performance>40\%) which is equal to now having any filter (all the population meet these criteria). It is noteworthy to mention that this experiment tests our proposed system in one of the populations simulated in the previous section. Based on our simulation and the results shown in \ref{fig:simulation performance difference}, we did not expect to obtain a better performance of CoBaCS in all the cases. Still, we wanted to test if the performance of CobaCS significantly exceeds ReBaCS.

In \cite{fiedler2019metacognition}, it is mentioned that the subjects' confidence is related to their response time; greater response time leads to less confidence. If this is true, we can measure the response time and not ask the users to rate their confidence. During our task, we measured the users' response time. According to the gathered data, in the memory task, this hypothesis is true (correlation coefficient is negative, p-value<0.05 for 71 users). On the other hand, in the tweet task, the correlation coefficient for just 42 users out of 86 is negative. This can be because the length of tweets is different, and each person has a different reading speed. Therefore, the findings in \cite{fiedler2019metacognition} cannot be applied to all the tasks. 

\begin{figure}
	\begin{center}
		\includegraphics[width=0.5\textwidth]{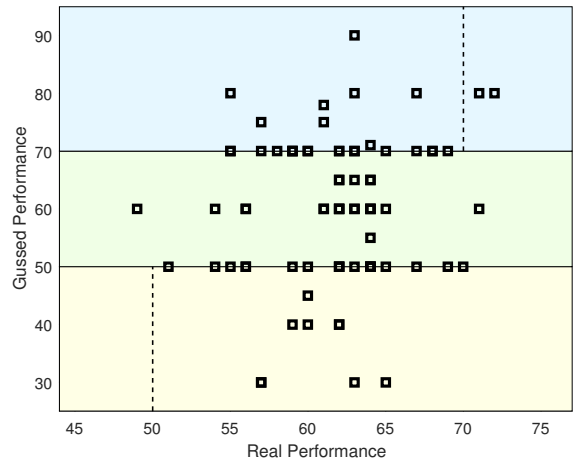}
	\end{center}
	\caption{DK effect: users who guessed their performance was below 50\%, their actual performance was greater than 50\%. Also, users who reported their performance greater than 70\% have actually performed with less than 70\% accuracy in most cases.}
	\label{fig: DK effect}
\end{figure}

As we mentioned in \ref{literature review}, the DK effect (Dunning-Kruger effect) refers to the cognitive bias in which worthy person rates her or his performance low, and the person with actual low performance reports that she or he had done the task properly. Also, in \cite{gadiraju2017using} is mentioned that using weighted majority voting in a population where the DK effect exists leads to more error compared with a majority voting-based system. To test whether this finding can be applied to our system, we asked users to rate their performance at the end of tweet tasks. As Figure \ref{fig: DK effect} shows, in our population, the DK effect exists, but the CoBaCS performance was significantly better than the ReBaCS. 

\section{Limitations and Scope}
This study has several limitations. First, the sample size was relatively small, consisting of 86 participants, which may limit the generalizability of the findings. Additionally, the participants were primarily from a specific geographic region, potentially introducing cultural or regional biases. Furthermore, there may be concerns regarding whether the tasks utilized in this study, such as the pre-tweet and memory tasks, comprehensively capture the full spectrum of cognitive abilities pertinent to all types of crowdsourcing platforms. However, our findings indicate that metacognition is not dependent on the specific tasks performed.

The scope of this study is focused on investigating the design and evaluation of crowdsourcing platforms based on users' confidence judgments. The primary aim was to enhance the recruitment workflow of these platforms by leveraging cognitive assessments. This research serves as an initial exploration into the relationship between user confidence and task performance within crowdsourcing environments.

\section{Conclusions}
In this paper, we first studied crowdsourcing systems. Then, we proposed our method, CoBaCS, which asks the users for not only their responses but also their confidence. Furthermore, this system did not include heavy mathematical calculations and can be implicated and run easily in online and offline crowdsourcing systems. Based on the research done in the field of meta-cognition, we measured the meta-cognition ability of the users and then used this measurement to select proper users. According to our findings, both mathematically and via a real-world experiment, our proposed system outperform the most popular crowdsourcing system (ReBaCS). As a result of our simulation, when the task is hard, and there are not enough experts present in the system, using CoBaCS is a wiser choice. Also, because the meta-cognition ability is not task-dependent and can be measured once a user is asked a variety type of questions, one can  use CoBaCS and filter users with low meta-cognition.

\bibliography{main}

\section{Data Availability Statement}
The datasets gathered and used during the current study will be available from the corresponding author (vmansouri@ut.ac.ir) on reasonable request.

\end{document}